# WRF-Chem optimization for the estimation of Etna volcanic ash fallout

Eleonora Brega[1], Maria Teresa Caccamo[2], Giuseppe Castorina[2], Salvatore Magazù[2,*], Mauro Morichetti[3], Gianmarco Munaò[2], Giorgio Passerini[1], Umberto Rizza[3]

[1] Department of Industrial Engineering and Mathematics Sciences, Università Politecnica delle Marche, Ancona, 60131, Italy

[2] Department of Mathematical and Informatics Sciences, Physical Sciences and Earth Sciences (MIFT), University of Messina, Viale F. Stagno D'Alcontres 31, 98166 Messina, Italy

[3] Institute of Atmospheric Sciences and Climate (ISAC) National Research Council (CNR), Unit of Lecce, Lecce, 73100, Italy

[*] Corresponding author: smagazu@unime.it

## Abstract

The main aim of the present work is to improve the quality of the Etna volcanic ash fallout forecasts through the optimization of an integrated simulation system based on the Weather Research and Forecasting (WRF) model coupled with the WRF chemistry (WRF–Chem) module. The proposed approach constitutes the first systematic application of the WRF-Chem based protocol to a specific Etna volcanic eruption, with possible effects involving the whole Mediterranean area. On that score, the attention has been focused on the eruption event, recorded on December 2015 from 3rd to 7th, which led to the closure of the Catania International Airport. Quantitative meteorological forecasts, analyses of the Etna volcanic ash transport and estimates of the ash ground deposition have been performed. In order to test the performance of the employed approach, the model outputs have been compared with data provided by satellite sensors, Doppler radars and weather stations. As a result, it emerges that, as far as the selected eruption event is concerned, the WRF-Chem model correctly reproduces the distribution of SO2 and of volcanic ash. Therefore, this model may provide a valuable support both to airport managements and to local stakeholders including public administrations.

**Keywords:**

# 1. Introduction

Airport security is constantly under threat due to extreme weather events, such as strong wind shear conditions and heavy rainfall, as well as due to natural hazards, like dust intrusions from desert and volcanic fallout. These latter may cause interruptions to flight operations and damage to the ground infrastructures (Wilson et al., 2014). As a consequence, flight cancellations and airport closures may occur, so generating in turn economic losses. In addition, the ash plume can cause serious damages to aircrafts during their flights. For example, it can be mentioned the emblematic case occurred on June 24[th] 1982. The British Airways Boeing 747, with 263 people on board, crossed a cloud of volcanic ash expelled during the eruption of the volcano Galunggung (Indonesia). This event, never recorded before, caused the shutdown of the four plane-engines (Hanstrum and Watson, 1983). In this circumstance, since the ash cloud was dry, the weather radars did not detect it; in fact, the radars were designed only to identify clouds on the basis of the humidity content inside them. The siliceous ashes also caused a sandblasting on the windshield and on the landing lights. Finally, they blocked the engines because, due to the high temperature inside the turbines, the ashes melted and settled on the surfaces of the vanes, causing a reduced air flow which, in turn, stopped the combustion inside the thrusters. An almost identical case occurred on December 15th, 1989, when a KLM Boeing 747, flying from Amsterdam to Anchorage, crossed the plume of the eruption of Mount Redoubt, undergoing the shutdown of all four engines due to the compressor stall. Again, once they got out of the ash cloud, the crew managed to restart the engines and landed safely in Anchorage.

The particulate expelled by volcanoes, consisting of small particles (typically ranging from few millimeters to a few micrometers), remains suspended in the air for usually long times, during which the particles retain their identity even if they are involved in physical-chemical processes in the atmosphere. For instance, suspended particles act as condensation nuclei for water droplets, favoring the nucleation process, i.e. the formation of hazes and clouds, and the occurrence of extreme weather phenomena, and acid rains which cause erosion and corrosion effects of materials (Curtius, 2006; Castorina et al., 2018). Based on the nature and size of the particles, one can distinguish between: aerosols (suspensions of particles with a diameter less than 1 μm), mists (formed by droplets with a diameter less than 2 μm), fumes (consisting of solid particles with diameter less than 1 μm and generally released by chemical and metallurgical processes), smokes (given by solid particles usually less than 2 μm in diameter and carried by gas mixtures), powders (solid particles with a diameter between 0.25 and 500 μm) and sands (solid particles with a diameter greater than 500 μm). A large amount of particles present in the atmosphere is constituted by fine particles with a diameter less than 2.5 μm, since the coarser ones, being heavier, are quickly deposited (Davies, 1974). Aerosols, in particular, are among the most widespread atmospheric components, and therefore they are the main responsible for air pollution. As a consequence, their increase contributes to the changes suffered by the Earth's atmosphere. This increase can be caused by human and natural sources. Among the latters, a fundamental role is played by the emissions caused by volcanic eruptions: during these processes, it is possible to observe the emission of aeriform, liquid (magma) and solid matter (pyroclastic materials that, depending on the size, are distinguished in ash, lapilli and volcanic bombs). In particular, the volcanic ash is usually classified according to its size (Mastin et al., 2009): the advantage of using such classifications is given by the possibility to get an estimate of the distance where the ash fallout is recorded and of the time employed for its gravitational settling. The aeriform component is given by the various released substances; these include water vapor, carbon dioxide, hydrogen, sulfur dioxide ($SO_2$), chlorine, nitrogen and various rare gases. The volcanic release in the troposphere of gaseous species composed of sulfur is one of the most important, and still unpredictable, causes of the variability of the natural climate: indeed, approximately 17% of the $SO_2$ in the atmosphere is attributable to volcanic eruptions (Diehl, et al., 2012). The $SO_2$ is well quantifiable by means of remote-sensing techniques performed with ultraviolet (UV) space-based instruments such as the Total Ozone Mapping Spectrometer (TOMS) (Schoeber, et al., 1993), the Ozone Monitoring Instrument (OMI) (Carn, et al., 2015) and the Ozone Mapping and Profiler Suite (OMPS) (Li, et al., 2017), supplemented by infrared measurements (IR) registered with HIRS, MODIS and AIRS (Prata and Bernardo, 2007). As an example of the deep influence of the volcanic emission on the global climate, it has been shown that the years immediately following highly explosive volcanic eruptions (where an enormous quantity of particulate matter is released into the atmosphere) are characterized by particularly harsh winters (McCormick et al., 1995; Robock, 2000). The origin of the decrease in the Earth's temperature after a substantial increase of the particulate in the atmosphere is due to a reflection effect of sunlight (short-wave); in any case this action is mitigated by the fact that the particles also reflect infrared radiation (long-wave) from the Earth (Frolicher et al., 2011).

With reference to both gas emission in the atmosphere and airport security, one of the most important volcanos whose activity needs to be continuously monitored is the Etna, which is placed in Sicily (Italy). In particular, the Etna volcano plays an important role being the largest source of emissions of gas and particles that are transported in the troposphere by winds for long distances covering a significant portion of the Mediterranean Sea (Wang et al., 2008). A study performed in the context of AEROCOM program (Diehl et al., 2012) reported that when considering the daily $SO_2$ emissions and plume heights for 1167 volcanoes from 1-1-1979 to 31-12-2009, Etna is the largest contributor with more than 45 Tg. This was also evidenced by the NASA Modelling, Analysis and Prediction (MAP) program (Tanguy et al., 2007). The Etna volcano is placed along the routes going from northern Europe to Africa, the Middle and Far East and Australia; furthermore, the connections between Sicily and the mainland take place mainly by air. In this framework, both the position of the Etna volcano in proximity to the airports (two airports within a distance of 25 miles) and the explosive activity of the volcano cause repeated flight cancellations and closures of the Catania International Airport. Furthermore, the eruptive activity affects numerous air routes, involving not only the Catania airport but also other regional airports like, Sigonella, Comiso, Reggio Calabria, Palermo and Trapani (see Fig. 1). From these considerations, it clearly emerges the need to develop an adequate monitoring of the Etna volcanic emission, through both the use of sensors and the prediction of volcanic ash transport.

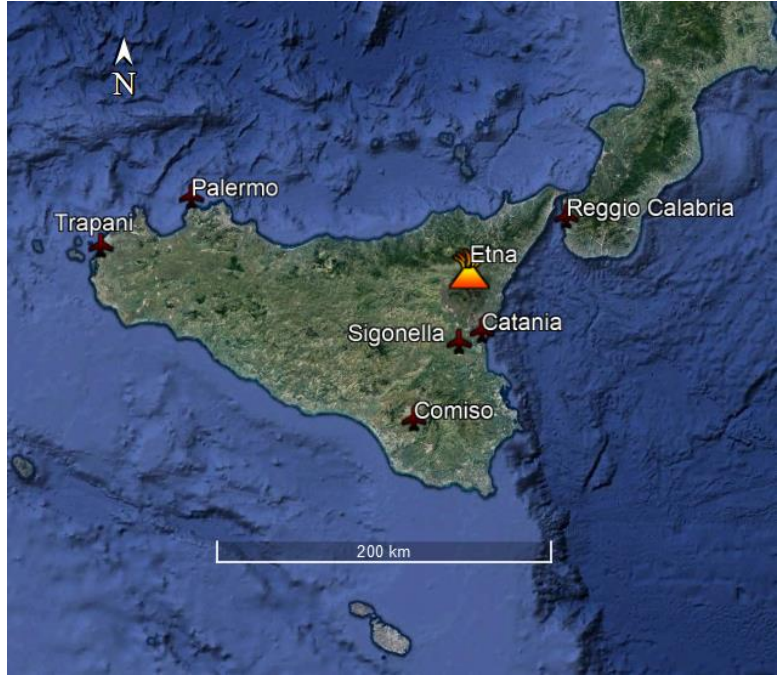

**FIG. 1**. Airports affected by the eruptive activity of the volcanic area of Etna.

In the present work we make use of the Weather Research and Forecasting model (Skamarock and Klemp, 2008) coupled with the chemistry module (WRF–Chem) (Grell et al, 2005; Misenis and Zhang, 2010), to describe in detail the significant Etna eruption occurred in December 2015, from 3[rd] to 7[th], which led to the closure of the Catania airport. This case study is thought as an application of the WRF-Chem model to investigate the effects of Etna eruptions on the air traffic of the involved airports. From a general point of view, the impact of the Etna emissions on the Mediterranean area has been largely documented by different techniques, including Doppler radar (Donnadieu et al., 2016, Freret-Lorgeril et al., 2018), Lidar (Favalli et al., 2009) and polar satellite (Andronico et al, 2009) monitoring: further studies concerning, among others, the tephra fallout (Scollo et al., 2007), the explosivity of basaltic magmas (Edwards et al., 2018), and the ash emission (Andronico et al, 2009b) of Etna volcano have been also performed.

The WRF-Chem package has been already applied to model emissions and transport of ash and $SO_2$ from volcanic eruptions all over the globe. In particular, Stuefer et al., (2013) made a significant development on the WRF-Chem software architecture exploring the functionality that permits simulating emissions, transport and settling of volcanic particles and gases, Webley et al., (2012) analyzed the Eyjafjallajökull 2010 eruption simulating with the WRF-Chem model the ash cloud dispersing toward mainland Europe. A coupling of aerosol clouds and radiative processes in the WRF-Chem model has been investigated by (Chapman et al., 2009; Zhang et al., 2010), while the effect of biomass burning has been taken into account by (Grell et al., 2011b). Further significant contributions to the development of the WRF-Chem package have been performed by (Misenis and Zhang, 2010; Shrivastava et al., 2011, Tuccella et al., 2012).

In the present study, in order to validate the proposed approach, we systematically compare the model predictions with observed data registered by satellite sensors, Doppler radar and mobile weather stations.

The paper is organized as it follows. In section 2, we describe in detail the considered case study, while the WRF-Chem model is introduced in Section 3. A semi-empirical description of eruption processes and sedimentation velocity is given in Section 4. The use of a pre-processor implemented in the WRF-Chem scheme is presented in Section 5, with results and discussion in Section 6. Finally, conclusions and perspectives are given in Section 7.

## 2. Case study: December 2015

In December 2015, from 3rd to 7th, at the Etna volcano, an explosive event occurred; it lasted about a week and was characterized by various episodes of different intensity. Moreover, these events had repercussions on the air traffic; in fact, the Catania International Airport reported numerous closures due to the volcanic ashed accumulated on the runway. In the weeks before the explosive event, a constant strombolian activity (explosions of varying intensity and frequency, with the casting of incandescent material sometimes accompanied by small amounts of volcanic ash) was observed inside the Voragine crater (often called "central"), one of the four summit craters of Etna. The strombolian activity intensified on the evening of December 2nd, culminating in a brief, but very violent, paroxysm on the night of December 3rd. Around 02:20 UTC the monitoring tools recorded the beginning of the eruption, which remained active until around 03:10 UTC. The eruption column exceeded 3 km in height, accompanied by volcanic explosions and lightning inside the ash cloud formed above the Voragine crater. The recorded data detected a signal/noise ratio of +18 dB and a longitudinal speed at the beam axis equal on average to 50 m s$^{-1}$, with peaks of 60 m s$^{-1}$. Moreover, the emissions of tephrites were in a range going from the 2.985 km up to 4.485 km, thus suggesting an intense expulsion of blocks and lapilli. The eruption of December 3rd was followed by other three similar paroxysmal episodes: two occurred on December 4th, respectively between 09:00 and 10:15 UTC and between 20:27 and 21:17 UTC, and the other one detected in the afternoon of December 5th between 14:40 and 16.20 UTC. Fig. 2 shows the intensity measurements of the volcanic activity (power), expressed in decibel-milliwatt (dBm) and taken from the database of the Physique du Globe de Clermont-Ferrand University (OPGC, http://wwwobs.univ-bpclermont.fr/SO/televolc/voldorad/).

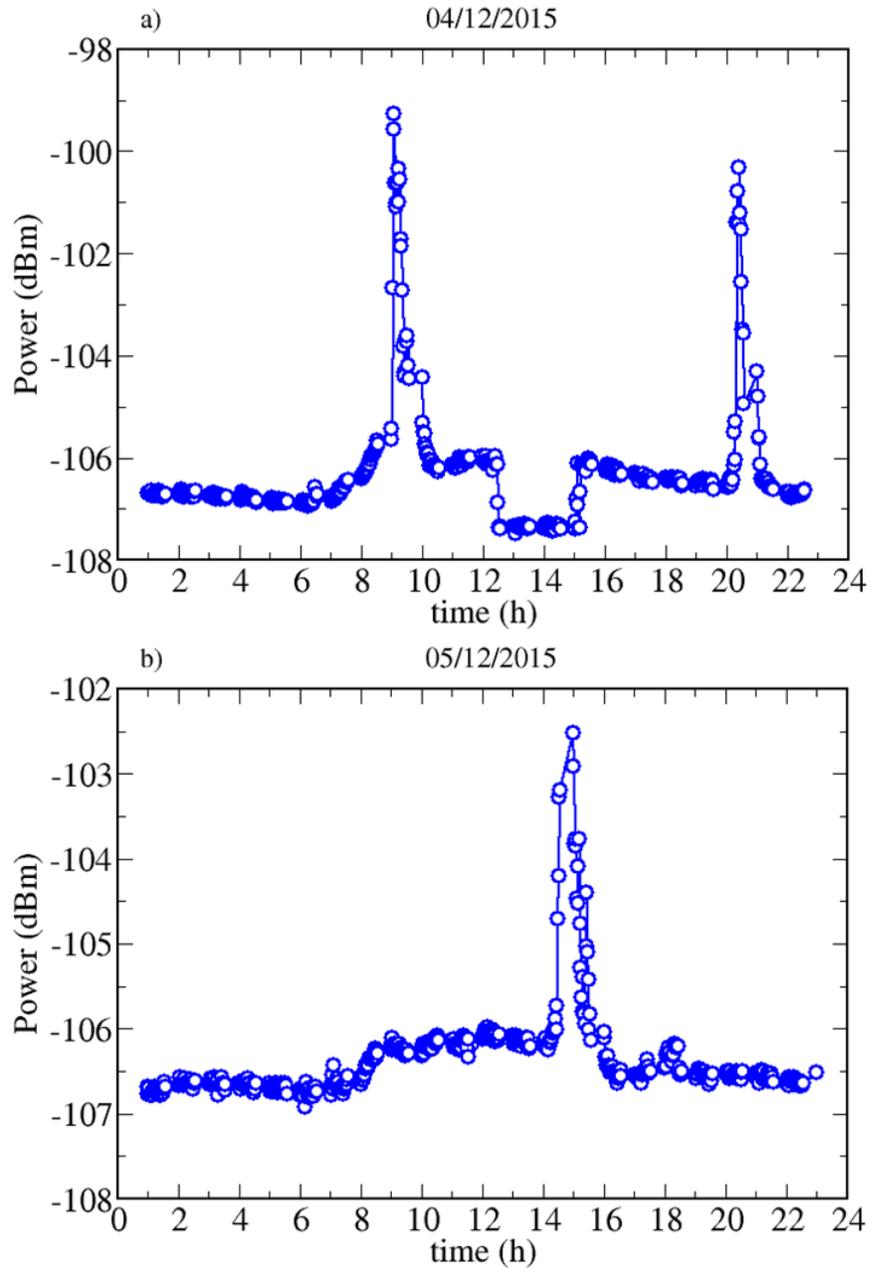

**FIG. 2.** Intensity of the Etna volcanic activity on December 4th (a) and December 5th (b). One observes the occurrence of 3 paroxysmal events: on December 4th, one between 09:00 and 10:15 UTC and the other one between 20:27 and 21:17 UTC; on December 5th, one between 14:40 and 16:20 UTC.

These three events were the most significant of the whole series of eruptions that affected Etna in this period of intense explosive activity. The eruption that occurred on the morning of December 4th was characterized by eruptive columns that reached about 7 km in height above the top of the volcano in about 15 minutes and at the same time frequent emissions of brown-greyish ash from a new crater were observed, along with high amplitude of volcanic tremor and strombolian explosions. The coarser pyroclastic material was deposited on the high south-western side of the volcano, at a quote of about 2 km, while the ash fallout occurred in its eastern sector, covering the Giarre-Zafferana Etnea area. A representative map showing the position of Giarre and Zafferana Etnea with respect to the Etna volcano is given in Fig. 3.

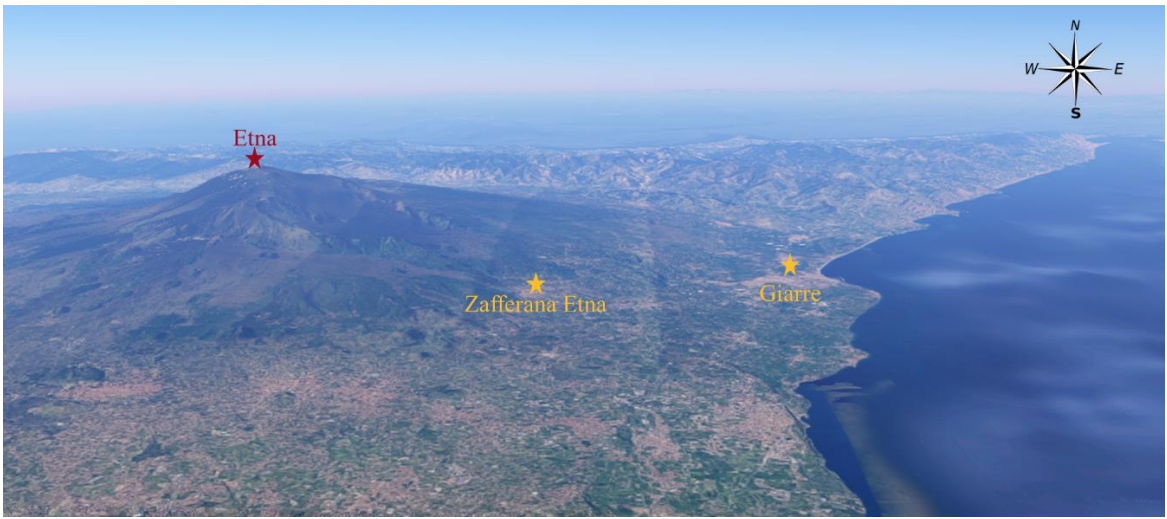

**FIG. 3.** Map showing the position of the Etna volcano along with Giarre and Zafferana Etnea.

The paroxysms that occurred on the evening of December 4th and in the afternoon of December 5th had almost the same characteristics as the previous ones, but with values of radial speed and power that have progressively decreased, leaving room for degassing phenomena in the late evening of December 5th. At the dawn of the following day the strombolian activity moved to the New Southeast Crater (NSEC), followed by landslides of unstable material (mixed with fragments of hot lava), which in a short time turned into burning avalanches. At the same time, a seismic swarm occurred in the north-eastern slope of Etna, probably connected with the volcano deflation due to the removal of a significant amount of magma: this deflation can be connected to the response of the tectonic structures in the eastern flank of the volcano. After about a week the situation became normal again and the explosive event finished.

## 3. The WRF – Chem model

The numerical simulation of the eruptive event was carried out by means of the WRF-Chem model (Grell et al., 2005) in its version 3.6.1 (https://ruc.noaa.gov/wrf/wrf-chem/). It consists of a type of "online" model as it allows the evaluation of the emission, transport, dispersion, transformation and sedimentation of all anthropogenic and natural pollutants. In the latter case, it is possible to consider both physical processes and chemical transformations of aerosol particles released during volcanic eruptions. Indeed, it is worth to note that, although weather conditions are the main factor which determines the air quality, they depend on the direct and indirect effects that chemical compounds have on the solar radiation and on the cloud microphysics. Most of the Volcanic Ash Transport and Dispersion (VATD) approaches are "offline" models, since they separately describe the physics and chemistry characterizing the dispersion of volcanic emissions into the atmosphere and generally, given their low computational cost, they are largely adopted by the operational forecasting centers. However, it is more realistic to privilege the "online" approach, since an inappropriate treatment of atmospheric processes could lead to incorrect answers concerning the deposition of the ash and the diffusion of aerosols (Grell and Baklanov, 2011).

The WRF-Chem is based on the WRF model (Skamaroch and Klemp, 2008; Castorina et al., 2017; Colombo et al., 2017; Powers et al., 2017; Castorina et al., 2018b) which supports both research applications and operational meteorological forecasting applications (Caccamo et al., 2017), including various options for dynamic cores and physical parameterizations (Castorina et al., 2019): therefore it can be used to simulate atmospheric conditions on a wide range of space and time scales. A chemical section that simulates tracer gases and particulate interactively with the meteorological fields making use of different photochemical treatments has been integrated by using the WRF-ARW numerical core (Skamarock et al., 2008). Among others, the WRF-Chem model can be used to forecast and simulate meteorological/climatic conditions on a regional and local scale, predicting the release and transport of pollutants and natural aerosols (Rizza et al., 2018), estimating the air quality and studying important processes for the global climate change (Zhang et al., 2010). The capability of the model to predict the transport and concentration of ash clouds and $SO_2$ depends on the initial information regarding volcanic emissions, such as the scale of the eruption (which includes the expelled mass), the height of the plume, the eruption rate, the date and duration of the event.

In our simulations the spectrum of the grain size distribution of the ash particles has been taken from Mastin et al., 2009 (hereinafter M09). Fig. 4 shows the numerical domain adopted in the present work: this domain includes a part of North Africa, Italy and a part of Balkan Europe with a 180 x 180 grid centered at a latitude of 37.74° and longitude of 15.18°. The horizontal space of the grid is 10 km in both directions with 40 vertical levels up to 50 hPa. The initial and boundary conditions were acquired by NCAR/NCEP Final Analysis (FNL by GPS) (ds083.2), with resolution of 1 degree, every 6 hours.

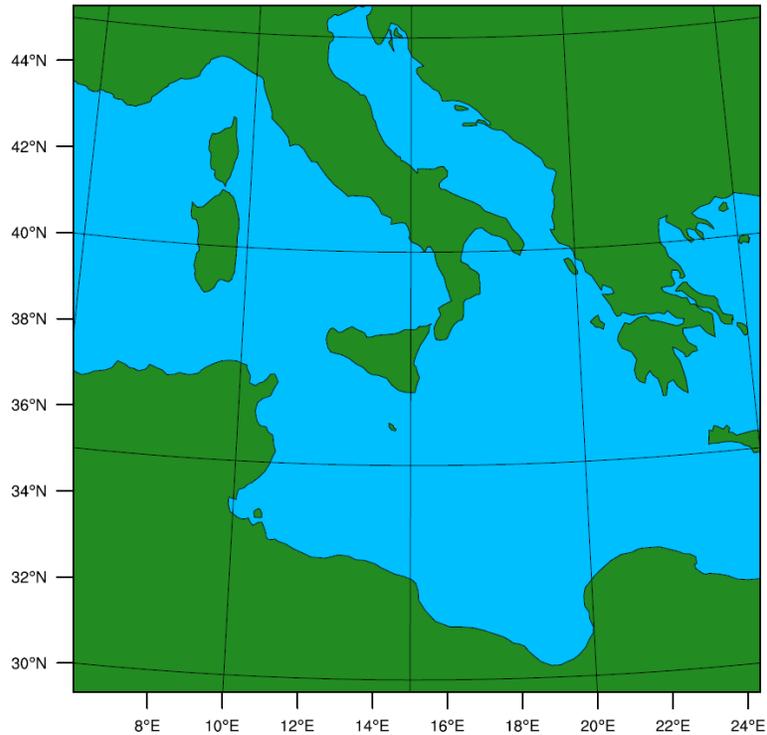

**Fig. 4.** Numerical Domain of the WRF-Chem model.

Table 1 shows the schemes selected for the setting of the parameterizations, specifically Mellor-Yamada-Nakanishi and Niino (MYNN - level 2.5) for the planetary boundary layer (Nakanishi and Niino, 2009), surface similarity (Paulson, 1970) and the surface model RUC (Rapid Update Cycle) (Benjamin et al., 2004). These schemes have been chosen to represent the physics of the surface layer and the parametrization of the Earth's surface. The Rapid Radiative Transfer Model (RRTMG) for both short-wave (ra_sw_physics = 4) and long-wave (ra_lw_physics = 4) radiation is used for the direct radiative effect of aerosol (Mlawer et al., 1997). The Purdue Lin scheme (mp_physics = 2) is used for microphysics processes. This configuration is compatible with short-wave radiative feedback (Bohren and Huffman,1983).

| | Option number | Namelist variable | Model |
|---|---|---|---|
| Land surface | 3 | sf_surface_physics | RUC model |
| PBL model | 5 | bl_pbl_physics | MYNN 2.5 level |
| Surface similarity | 1 | sf_sfclay_physics | MM5 Similarity Scheme |
| Microphysics | 2 | mp_physics | Purdue Lin |
| Short-wave radiation | 4 | ra_sw_physics | RRTMG |
| Long-wave radiation | 4 | ra_lw_physics | RRTMG |

**TABLE 1.** List of the variables that constitute the physical parameterization of the WRF–Chem model.

The implementations of the other parameters follow the guidelines of M09 (such as the mass of dust, set to 1.7 x $10^9$ kg), or are based on estimates derived from the literature (as for the mass of $SO_2$ set to 1.0 x$10^9$ kg) (Schnetzler et al., 1997). These parameters are reported in Table 2: for the injection heights and duration, the observations taken from the cameras and radar Doppler VOLDORAD 2B (http://voldorad.opgc.fr/home.php) have been adopted. Such a radar is placed at approximately 3km from the top of the volcano and has been made available by the Observatoire de Physique do Globe de Clermont-Ferrand (OPGC) in collaboration with the Istituto Nazionale di Geofisica e Vulcanologia (INGV) of Catania.

| Case | Eruption begin | Duration (sec) | Injection height (m) | Ash mass (kg) | SO₂ mass (kg) | Chem opt |
|------|---------------|----------------|----------------------|---------------|---------------|----------|
| 01 | 2015/12/04 – 09:00 | 3600 | 7000 | $1.7 \times 10^9$ | $1.0 \times 10^9$ | 402 |
| 02 | 2015/12/04 – 20:30 | 3600 | 7000 | $1.7 \times 10^9$ | $1.0 \times 10^9$ | 402 |
| 03 | 2015/12/05 – 14:55 | 3600 | 7000 | $1.7 \times 10^9$ | $1.0 \times 10^9$ | 402 |

TABLE 2. Volcanic setup of the WRF-Chem model.

## 4. Semi-empirical description of eruption processes and sedimentation velocity

The classification drawn up by M09 is based on the study of well-documented past eruptions and includes eleven types of eruptions. If new observations reveal data different from the predefined ones, one needs to find the type of eruption closest to the one under study and use its parameters. In the M09 classification concerning the granulometric size spectrum, the ash particles are divided into 10 bins (with diameters ranging from 2 mm to less than 3.9 μm) each one with the corresponding percentage of the mass fraction depending on the type Eruption Source Parameters (ESP). The latter are needed to predict the transport and dispersion of volcanic ash clouds during eruptions, with the aim of identifying risks for the air navigation. For each ash-bin (Table 4) it is possible to define the aerodynamic radius as half of the arithmetic mean between the limits of the diameters of each ash-bin. The aerodynamic radius can be used, in turn, to calculate the sedimentation velocity: the latter is a useful parameter in the context of the volcanic ash fallout, since it provides indication on the velocity experimented by a particle of a given size when it reaches the ground. In the present work a simple expression of the sedimentation velocity is presented, by following the procedure presented in (Pruppacher and Klett, 2010). At this purposes, we first need to define the diffusion coefficient $D$ through the relation:

$$D = \frac{kT(1 + \alpha N_{Kn})}{6\pi\eta_a r} \tag{1}$$

where $N_{Kn}$ is the Knudsen number, $\eta_a$ is the air viscosity, $k$ is the Boltzmann constant and $T$ is the temperature. The factor inside the parenthesis, is named Cunningham factor, and, besides $N_{kn}$, it depends on the numerical parameter $\alpha$, defined as (Pruppacher and Klett, 2010):

$$\alpha = A + B exp\left(\frac{-C}{N_{Kn}}\right). \tag{2}$$

In this equation, the numerical parameters A, B and C have been empirically set as A = 1.257, B = 0.400, C = 1.10. The Knudsen number is defined as:

$$N_{Kn} = \frac{kT}{\sqrt{2}\pi r^2 PL} \tag{3}$$

where $P$ is the atmospheric pressure and $L$ is a representative length scale of the system.

Once defined the diffusion coefficient, the drift velocity can be calculated as:

$$\boldsymbol{v}_d \equiv \frac{D}{kT} \boldsymbol{F}_{ext} \tag{4}$$

where, for spherical particles of density $\rho$ and subjected to gravitational force, the external force $\boldsymbol{F}_{ext}$ is defined as:

$$\boldsymbol{F}_{ext} = \frac{4\pi}{3} r^3 (\rho - \rho_a)\boldsymbol{g} \tag{5}$$

where $\rho_a$ is the air density and $\boldsymbol{g}$ is the gravity acceleration. The sedimentation velocity $v_s$ is finally defined as:

$$v_s \equiv |\boldsymbol{v_d}| = \frac{2(1+\alpha N_{Kn})r^2(\rho-\rho_a)g}{9\eta_a} \qquad (6)$$

Therefore, the sedimentation velocity can be easily calculated as a function of the aerodynamic radius of the ash particles.

As far as Etna is concerned, the M09 classification assigned the class M1 to its eruptions, which are mainly of an effusive type, *i.e.* they are characterized by basaltic dense-lava fountains, with relatively high temperatures (> 1000 ° C). In Table 3 it is reported the particle size distribution assigned to the emitted particles diameters in the range between 1000 μm and 31.25 μm. According to the classification reported in the table, 4% of particles have a diameter between 0.5 and 1 mm (vash_2), 10% between 0.25 and 0.5 mm (vash_3), 50% between 125 and 250 μm (vash_4), 34% between 62.50 and 125 μm (vash_5) and 2% between 31.25 and 62.50 μm (vash_6). No particles with diameter less that 31.25 μm are considered (Stuefer et al., 2013).

| Ash bin | Mm | % |
|---------|-----|---|
| 1 | [1000 - 2000] | 0 |
| 2 | [500 - 1000] | 4 |
| 3 | [250 - 500] | 10 |
| 4 | [125 - 250] | 50 |
| 5 | [62.50 - 125] | 34 |
| 6 | [31.25 - 62.50] | 2 |
| 7 | [15.62 - 31.25] | 0 |
| 8 | [7.81 - 15.62] | 0 |
| 9 | [3.91 - 7.81] | 0 |
| 10 | [0 - 3.91] | 0 |

**TABLE 3.** Particle size distribution characterizing the particles emitted by the Etna volcano, according to the study by M09.

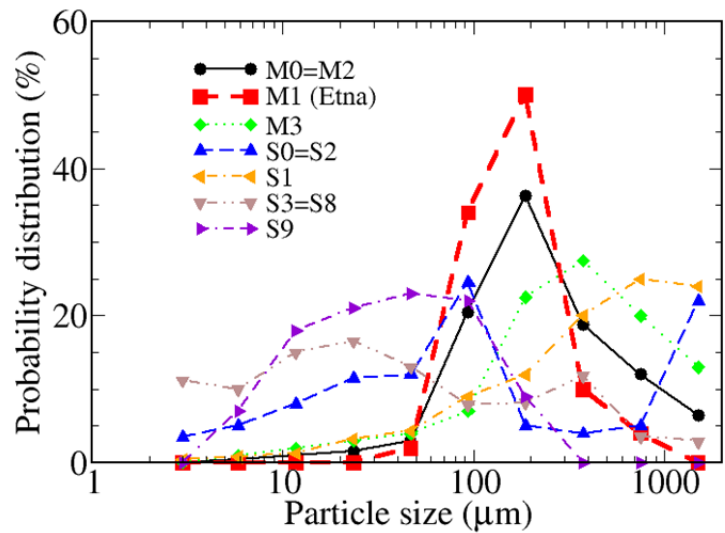

**Fig. 5.** Probability distribution in a semi-logarithmic scale of particle sizes of volcanic ash emitted by Etna (M1) compared with other class of eruptions.

In Fig. 5 the distributions of volcanic ash emitted by Etna (M1) are compared with those observed in different classes of eruptions, according to the database by M09. It emerges that the distribution of Etna ash shows a well-defined peak at a particle size of 175 µm, with about 50% of emitted particles having this size. In addition, it is worth noting that this is the less polydisperse distribution, with only five values other than zero. These findings suggest that this particular class of eruptions may attain very high temperatures, in order to generate a huge amount of particles of relatively large size with a low degree of polydipersity.

Finally, once known the typical size of volcanic ash emitted by Etna, it is possible to estimate the sedimentation velocity by using Eq. 6. The results are reported in Fig. 6: upon increasing the particle size, the velocity increases by following a parabolic trend in agreement with the $r$-dependence of $v_s$ shown in Eq. 6. In the inset we report the probability distribution of the sedimentation velocity, finding that the most probable value (corresponding to a particle size of 175 µm) is around 11 m s$^{-1}$.

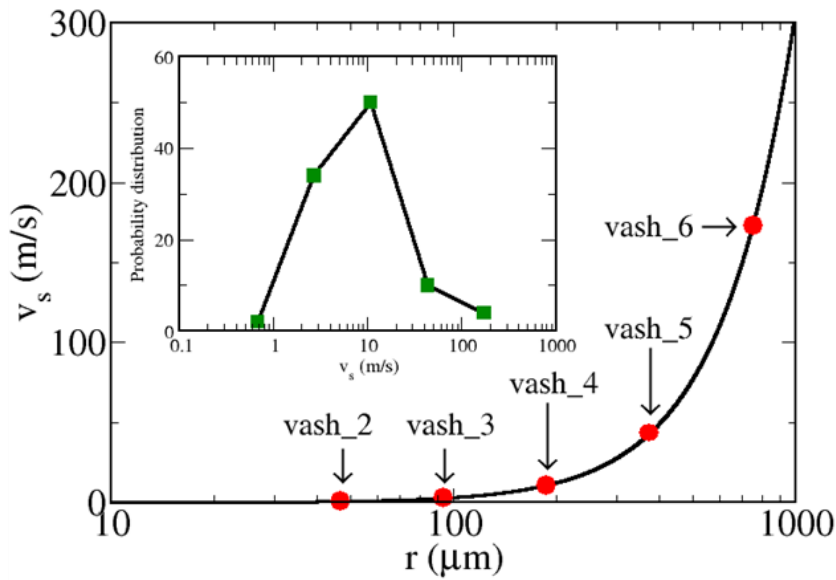

**Fig. 6.** Behavior of sedimentation velocity as a function of the particle size. The different vash sizes are explicitly reported. Inset: probability distribution of the sedimentation velocity.

## 5. The PREP-CHEM-SOURCES

The input parameters of the WRF-Chem model are determined by using a software pre-processor called PREP-CHEM-SRC (Freitas et al., 2011). The software is written in a hybrid Fortran-90 and C language and is able to read the global emissions of the most important anthropogenic pollutants, which are generally in HDF (Hierarchical Data Format) and NetCDF (Network Common Data Format). In this article the Version 1.5 of the PREP-CHEM-SRC has been employed. The desired grid configuration, the emission inventories and the additional information are defined in a file called "prep-chem-src.inp", while the outgoing emissions are provided in intermediate binary files, separated by type of emission. The convert_emiss program (included in the WRF-Chem package) converts them into NetCDF, or in input data files for the WRF-Chem, calculating the vertical mass distribution and ash emissions for each size before filling the input matrices, which will be analysed by WRF-Chem. The latter recalculates the emissions based on a new eruption height and vertical mass distribution (supplied as an input parameter) or uses the prescribed ash emissions (Stuefer et al., 2013).

The PREP-CHEM-SRC provides the position of the volcano in the numerical domain considering the nearest grid cell and the emission input parameters such as the height of the eruption column, the mass eruption rate and duration, if no other observation is available. This information is used within the WRF-Chem to determine the vertical distribution of the ejected material. In addition to the ash, during a volcanic eruption a considerable amount of $SO_2$ is also emitted which, once in the atmosphere, oxidizes and transforms into sulfuric acid ($H_2SO_4$); the latter condenses into sulfate aerosols, characterized by a residence time in the atmosphere proportional to the gases containing sulfur present in the volcanic plume. Unlike the ash that settles within few days, $SO_2$ can last up to several months (McCormick et al. 1995).

The data provided by the international program Aerosol Comparisons between Observations and Models (AeroCom, started in 2002 with the target to reduce the uncertainty on the impact of aerosols on the climate system) contain the volcanic emissions of $SO_2$ and other variables for the time interval going from January 1979 to December 2009 for all the volcanoes listed in the Global Volcanism Program database provided by the Smithsonian Institution. In particular, there is a file for each year containing the number of events occurred, along with the name of the volcano, the date, the height above the average sea level, the height of the plume, the longitude, the latitude and the daily emission rate of $SO_2$, distinguishing between volcanic eruptive and non-eruptive emissions.

The pre-processing tool, as in the case of volcanic ash, places the $SO_2$ emissions in the grid point of the WRF-Chem domain that surrounds the geographical position of the investigated volcano. The total emission is calculated by adding the emissions of all the volcanoes inside the cell and is expressed in Kg $km^{-2}$ $hr^{-1}$. However, it is important to point out that this value is often only approximated, in consideration of the difficulty usually met in accurately estimating the correct total amount of $SO_2$ emitted during a volcanic eruption.

## 6. Results and discussion

### 6.1 Geopotential

In Fig. 7 (a-d) the comparison between the geopotential height at an atmospheric pressure of 500 hPa predicted by the WRF-Chem (left panels)  and data downloaded from ECMWF/ERA5 Reanalysis (right panels) is reported. During December 4[th] (fig.7a) ERA5-reanalysis (right panel) shows a pressure minimum localized between the southern coast of Sardinia and northern coast of Tunisia that foster a south-east circulation in correspondence of Etna. The following day, December 5[th] (fig.7b, right panel) ERA5-reanalysis depicts that the pressure minimum is further dislocated toward south entering the African continent in correspondence of Tunisia/Libyan border. The consequent circulation at Etna is rotated in direction North-West. On December 6[th], again considering ERA5-reanalysis (fig.7c, right panel) it is evident a deepening of the low pressure between Balearic Islands and Sardinia producing northward wind in correspondence of Etna. Finally, on December 7[th], when the eruptive intensity started to decrease, ERA5-reanalysis (fig.7d, right panel) show that the cyclonic system dislocated in the north-west direction toward the southern coasts of French, inducing a northern flow in correspondence of Etna. It is evident the close reproduction of the ERA5 (fig.7, right panels) geopotential at 500 hPa by the model WRF-Chem (fig.7, left panels). This is quite important as the dispersion of eruptive plumes of ashes and gases (SO2) is strictly connected with wind speed and direction.

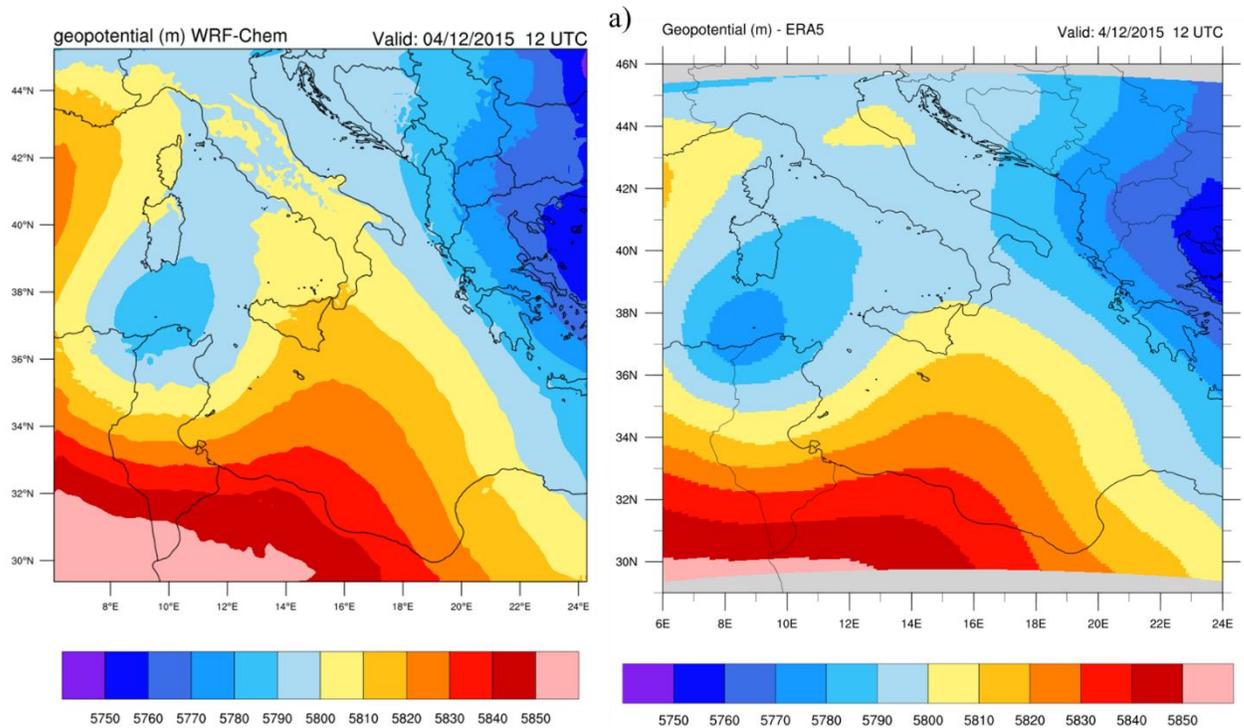

a)

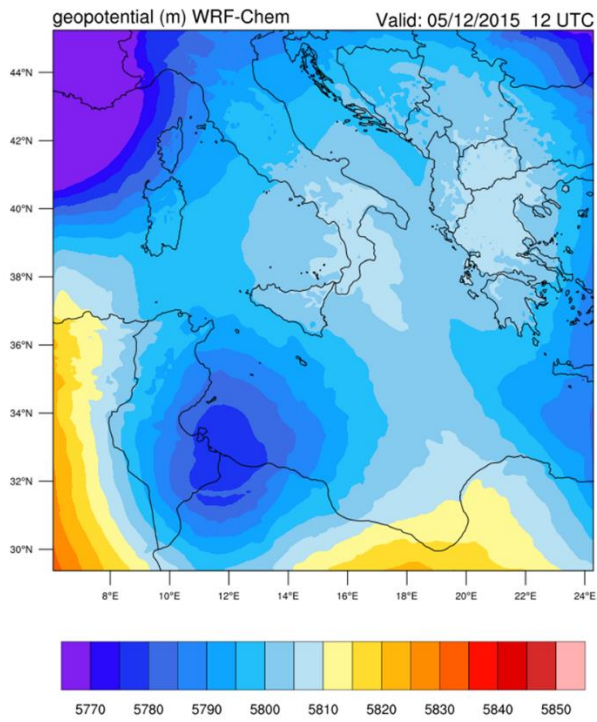

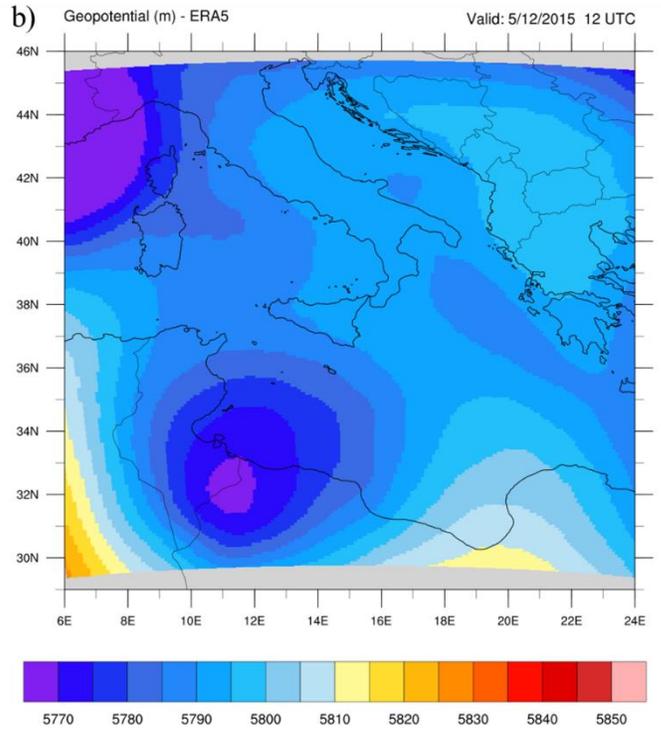

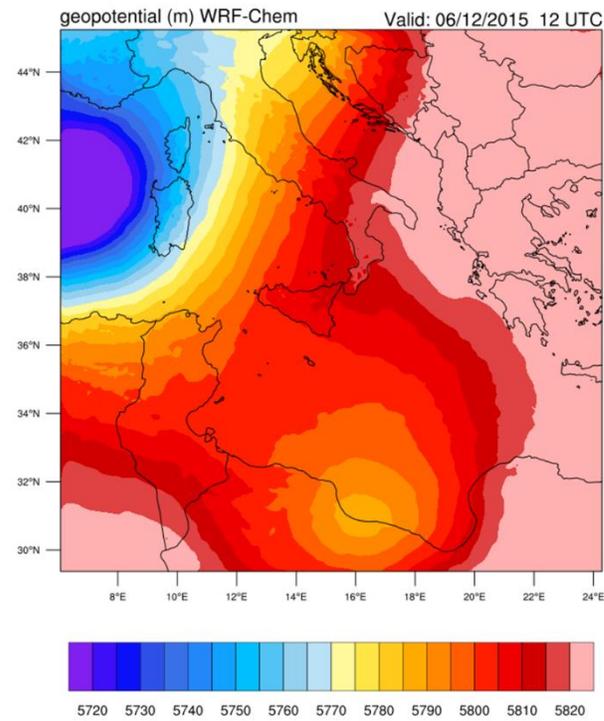

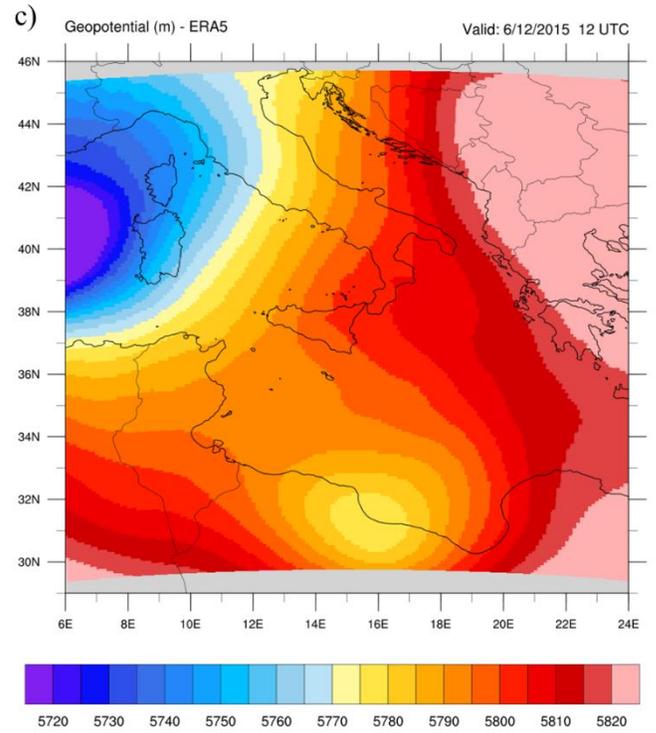

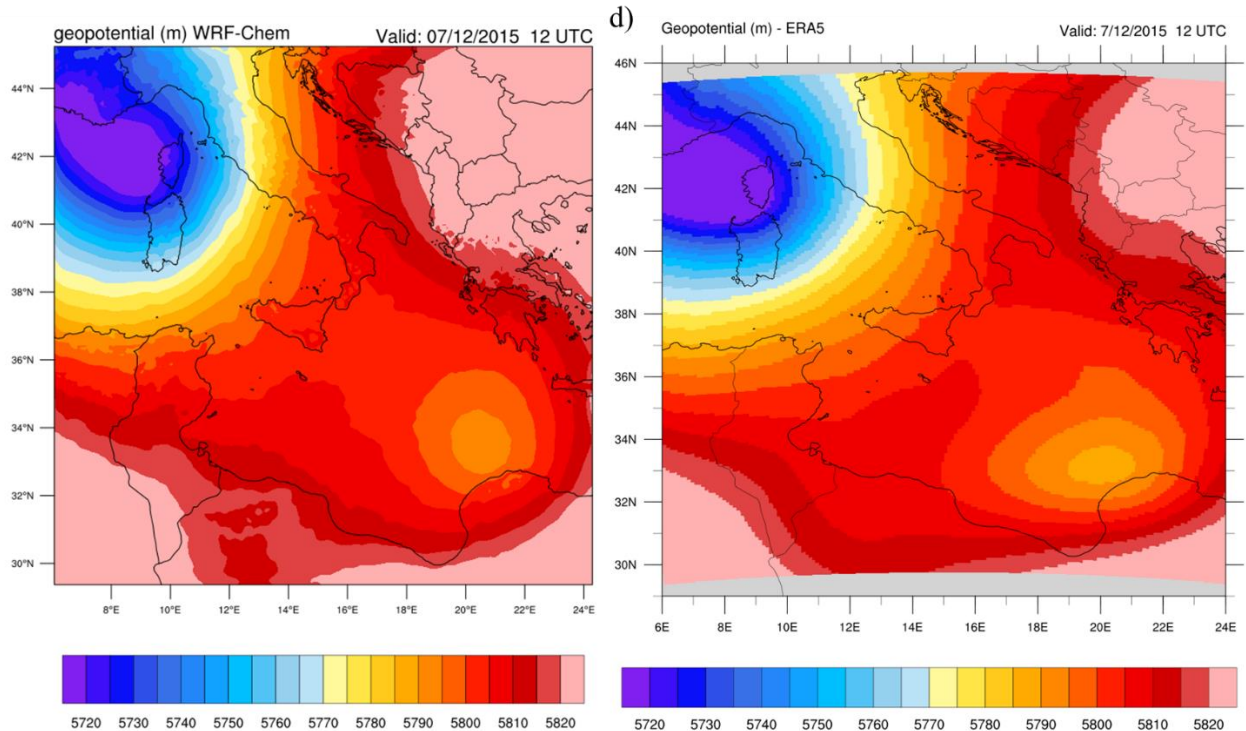

**FIG. 7** Geopotential height at an atmospheric pressure of 500 hPa checked on 4 days a) 4[th] December, b) 5[th] December, c) December 6[th] d) December 7[th]. Left panels: output from WRF-Chem simulations. Right panels: observed data from ERA5 – ECMWF website (https://cds.climate.copernicus.eu/cdsapp#!/dataset/reanalysis-era5-pressure-levels?tab=overview).

- **6.2 Transport of SO₂**

Sulfur dioxide ($SO_2$) is a colorless, irritating, non-flammable gas, very soluble in water and with a pungent smell. This compound is one of the most widespread, dangerous and studied atmospheric pollutants. It derives from the oxidation of sulfur during the combustion processes of substances that contain this element. Due to the characteristics of sulfur dioxide, it is interesting to analyze its distribution in the course of the studied event. At this purpose, the daily data relating to the columnar density of $SO_2$, expressed in Dobson Units (DU), have been downloaded from the NASA/Earthdata program

(https://earthdata.nasa.gov) through one of its dedicated web-portals (https://giovanni.gsfc.nasa.gov/giovanni/). The DU indicates the concentration of gas in a column of air, above a certain point on the Earth's surface. The daily $SO_2$ total column data have been provided by the EOS-Aura satellite (https://aura.gsfc.nasa.gov) via the sensor OMI (Ozone Monitoring Instrument) with a spatial resolution of 0.25 ° x 0.25 °. The $SO_2$ experimental data have been compared with the output of the WRF-Chem model: the latter have been first averaged over a single day and then column-integrated in every surface grid point, finally providing values expressed in µg m$^{-2}$. The conversion factor between DU and µg m$^{-2}$ is: 1 µg m$^{-2}$ = 3.5 x 10$^{-5}$ DU.

The WRF-chem simulations allow to analyze the spatial distribution of daily averaged sulfur dioxide emitted during the eruptive event under investigation. The scale of the daily average $SO_2$ concentrations shown in Fig. 8 identifies values ranging from 0 to 100000 µg m$^{-2}$ that correspond to [0-3.5] DU. In particular, on December 4$^{th}$ (fig.8a) there is an intense emission of $SO_2$ that spread out mainly to the south-east with a plume that crosses the Mediterranean until it almost reaches the coasts of Africa.

Between December 5$^{th}$ and 7$^{th}$ (figs. 8b, 8c and 8d) another intense flow of $SO_2$ from Etna is transported to the north, covering the entire Italian southern peninsula until the central Adriatic regions. In this context, it may be evidenced that the results of the simulations provide a spatial distribution of the daily averages $SO_2$ that reflects in a fairly precise way the synoptic analysis and the description of the event discussed above (see Fig. 7).

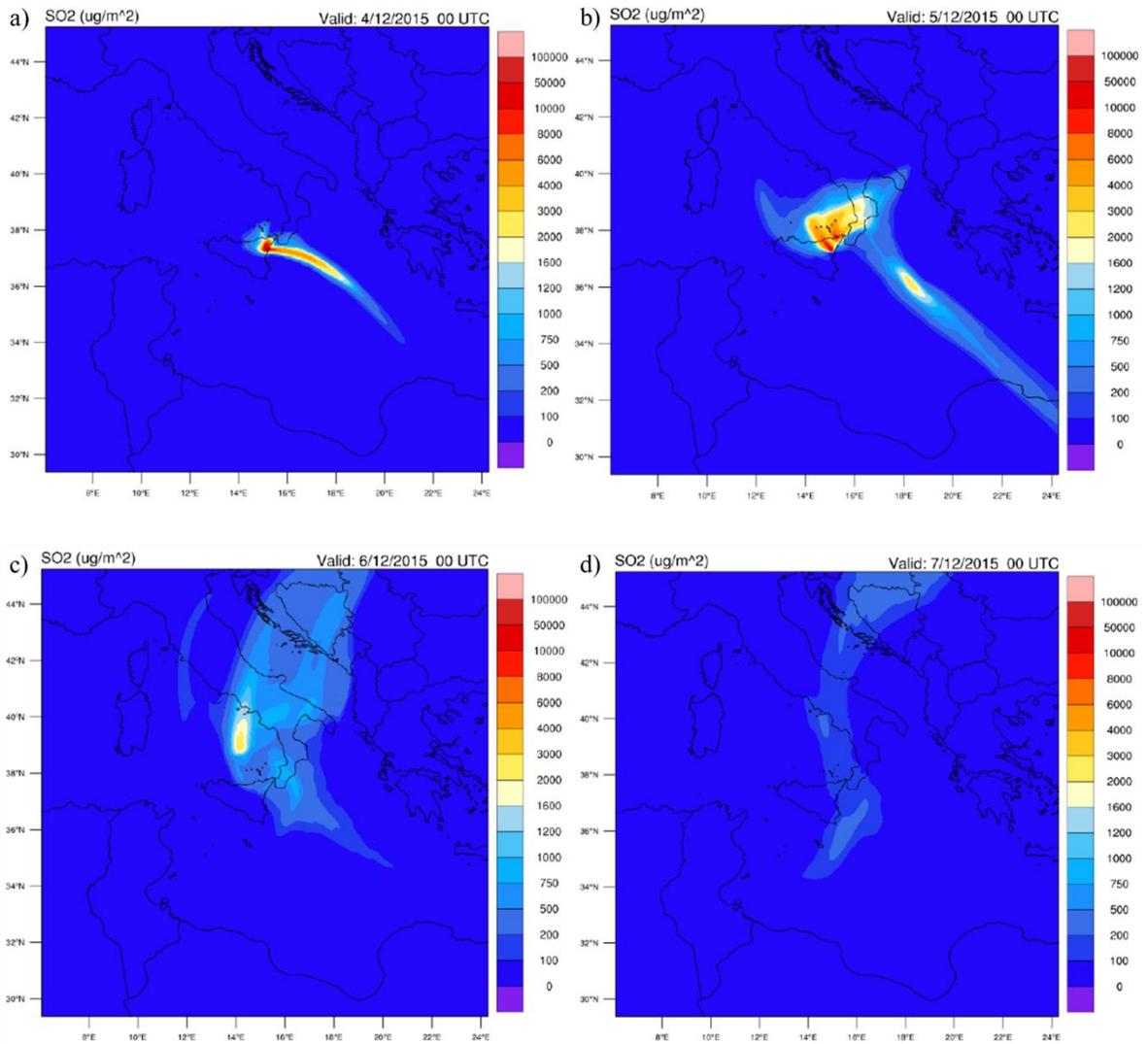

**FIG. 8.** Sequence of images illustrating the distribution of sulfur dioxide (SO2), obtained from simulations with the WRF-Chem model.

A comparison between the results obtained by using the WRF-Chem model and the daily data retrieved by the Aura satellite via the OMI sensor has been also performed. At 13:00 UTC of December 4th, immediately after the paroxysm, the OMI map reports approximately the same distribution of $SO_2$, characterized by a shift of the flow towards the south-east (Fig. 9a). According to the simulation results, the $SO_2$ concentration amounts to approximately 50000 μg m$^{-2}$ in proximity of the volcano. This value corresponds to 1.75 DU, close enough to the observed value of about 2 DU. Due to the lack of satellite data it is not possible to make a comparison on December 5$^{th}$. On December 6$^{th}$ (Fig. 9b), there is a good qualitative correspondence between the spatial distribution predicted by the model and that provided by the satellite, albeit the amount of SO$_2$ appears slightly underestimated by the simulations. In particular, it is possible to observe from both maps the transport of the SO$_2$ to the north, up to the Gargano and central Adriatic regions. Finally, with reference to December 7$^{th}$ (Fig. 9c) both maps show that sulfur dioxide, spread to central Italy, decreased in concentration, although it is not possible to explain the presence of an abundant quantity of SO$_2$ recorded by the OMI sensor between Sicily and Calabria. The origin of this discrepancy can be ascribed to the fact that in the WRF-Chem approach we simulate only the three main events (see Table 2), whereas puffs of erupted material can be expected along the entire duration of the event (degassing). Despite the discrepancy observed on the last day of the eruption, the overall distribution of SO$_2$ along the whole duration of the process appear fairly well reproduced by the WRF-Chem model.

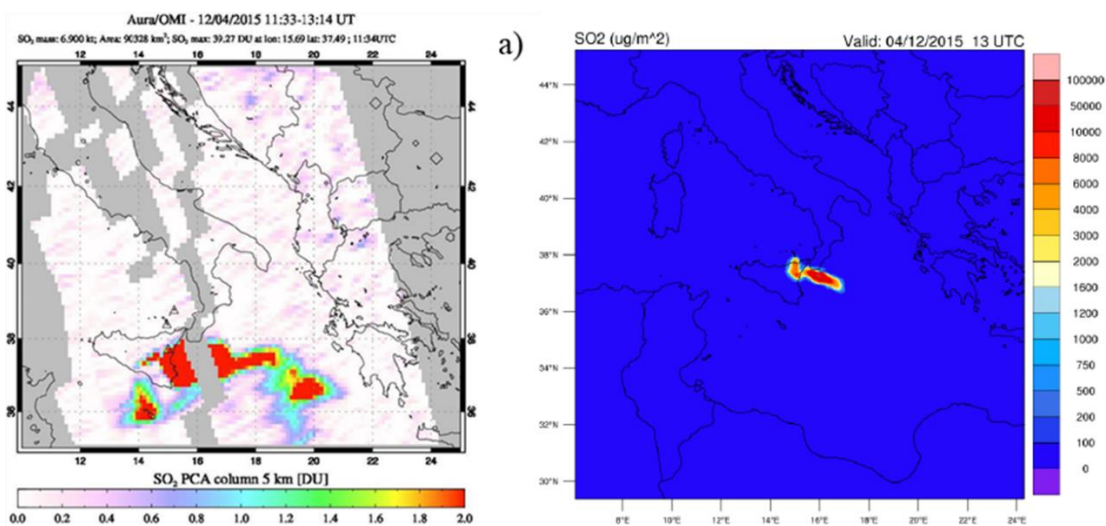

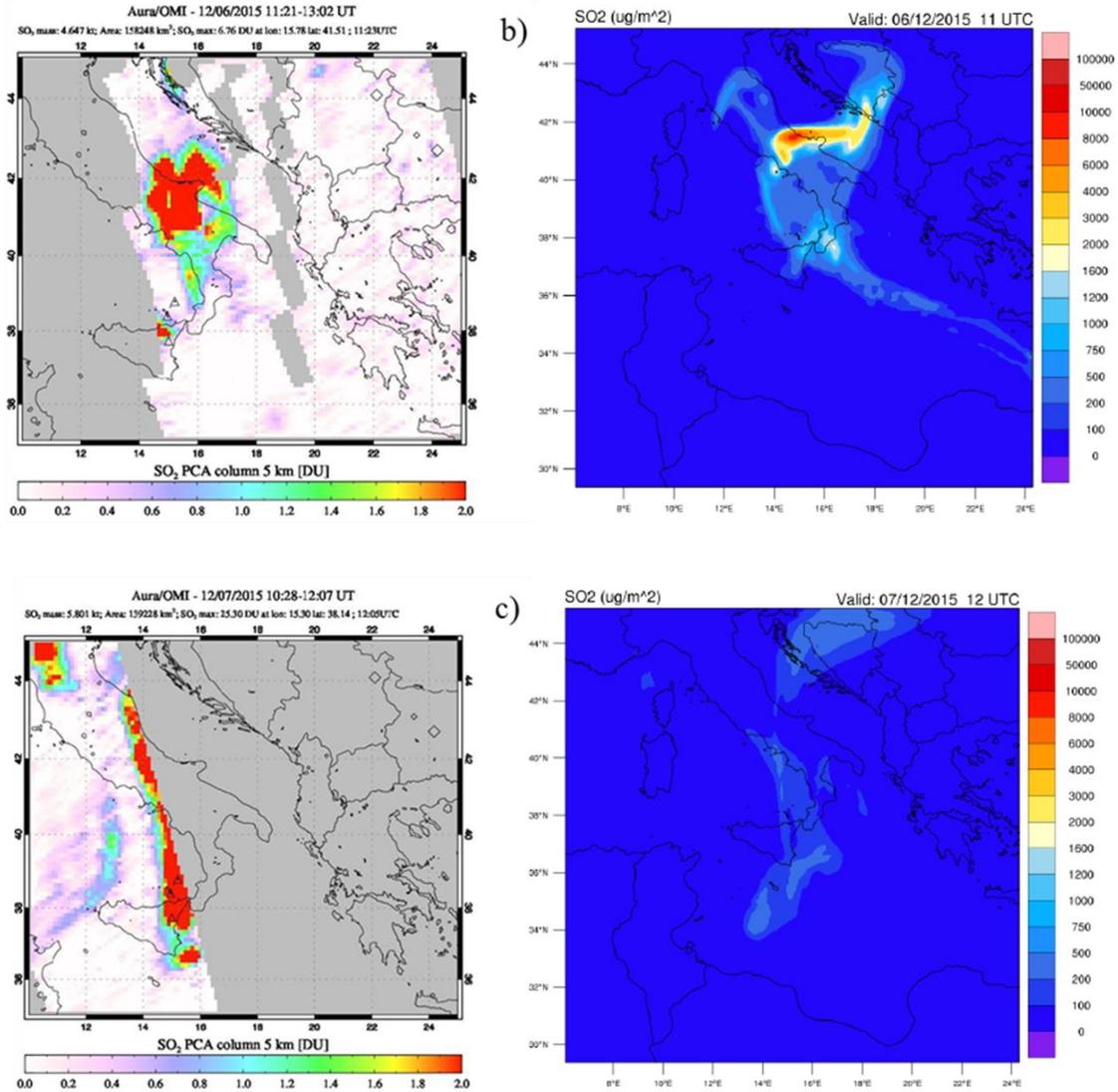

**FIG. 9.** Comparison between the results of the daily data acquired by the Aura satellite (left) and the simulations performed with the WRF-Chem model (right), relative to the transport of SO$_2$. The maps referring to the days of 4$^{th}$ December at 13:00 UTC (a), 6$^{th}$ December at 11:00 UTC (b) and 7$^{th}$ December at 12:00 UTC (c) were chosen.

- ## 6.3 Distribution of volcanic ash

In addition to the $SO_2$, in this study the transport of vash_4, vash_5 and vash_6 has been simulated for the time interval from December $4^{th}$ to $7^{th}$. As reported in Fig. 10, the average values of the different vash sizes range between $10^{-10}$ and $10^4$ µg m$^{-2}$. It is observed that the larger particles (vash_4), being heavier, have fallen back into the areas adjacent to the volcano, while the finer ones (vash_6), and therefore lighter, have been transported to greater distances, reaching central Italy and African coasts.

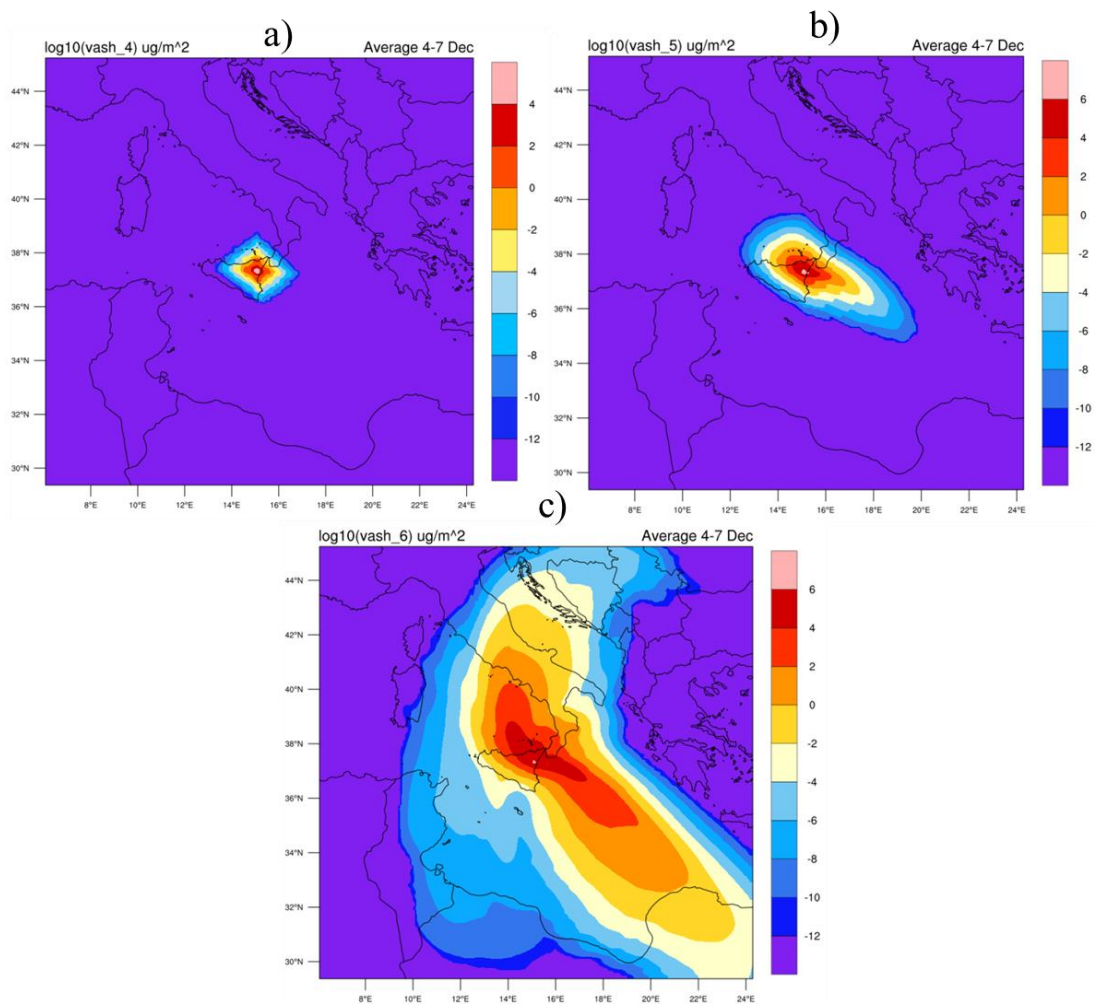

**FIG. 10**. Transportation of volcanic ash (a: vash_4, b: vash_5 and c: vash_6) in the period from $4^{th}$ to $7^{th}$ December.

In order to validate the model predictions, the simulation results have been compared with observed data characterized by the Aerosol Index (AI), explicitly developed to detect the presence of aerosols in the air. The aerosol measurements were carried out again by the OMI sensor on board the above described Nasa/Aura spacecraft. In Fig. 11 the area circumscribed by the dashed line highlights the regions affected by the dispersion of aerosols due to the eruption of Etna, showing how the suspension of particles is spread mainly over Sicily, the Mediterranean Sea and the Gargano, in agreement with the wind circulation derived from the synoptic analysis previously described. There are also high concentrations of aerosols (orange coloring) in the south-east part of Sicily, but since the wind mainly blew northwards during the paroxysmal episodes, this can be ascribed to dust from the Sahara. A qualitative agreement with the WRF-Chem results is clearly visible, albeit a more detailed comparison is not possible due to the different units adopted.

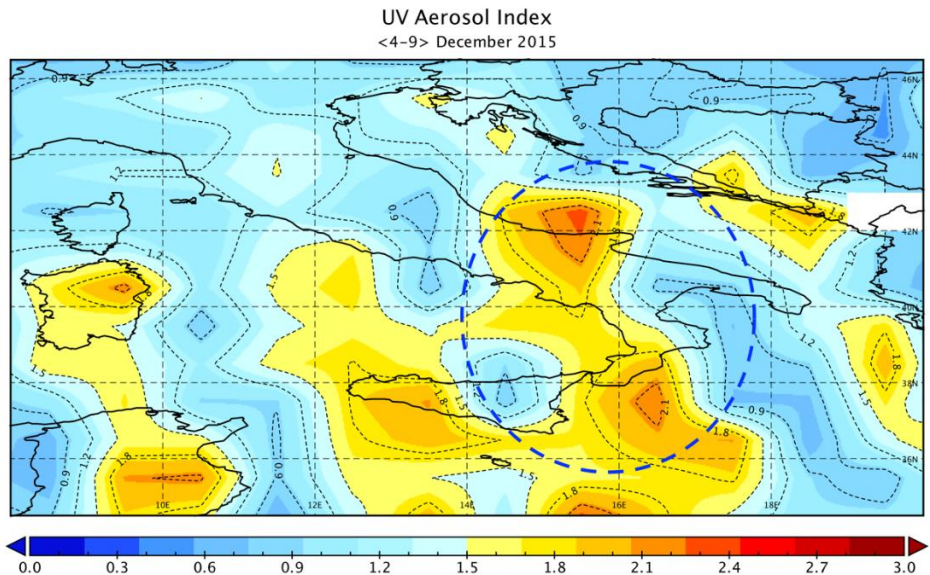

**FIG. 11**. Illustration of the average distribution of aerosols through the data acquired by the OMI sensor, in the period from 4th to 9th December 2015. The area delimited by the dashed blue line highlights the presence of volcanic aerosols in the domain of interest of this study.

Finally, a further analysis has been done considering only the volcanic ash with diameter between 15.62 and 31.25 μm (vash 7) that according to M09 distribution reported in table 3 should not be present for M1 class. This supplementary study aims to investigate the transport of small particles in the numerical domain. In Fig. 12 the WRF-Chem predictions on the transport of vash_7 is shown: as observed, in this case the finest volcanic particles can be found in the eastern regions of Europe, including Slovenia, Croatia, Bosnia and Herzegovina and Serbia. This may be investigated and verified by future studies by considering LIDAR measurements in these regions.

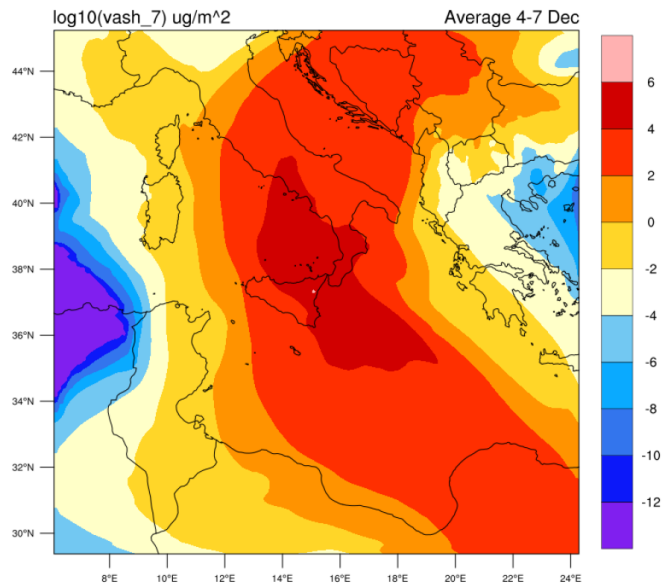

**FIG. 12**. Transportation of volcanic ash considering vash_7 (not included in the M1 classification) in the period from 4th to 7th December.

## 7. Conclusions

In this work we have presented the first application of the WRF-Chem model to systematically investigate the Etna volcanic eruption registered at the beginning of December 2015 and its role in affecting the security of the airports involved as well as the transportation of pollutants and volcanic ash in the Mediterranean area. As far as the case study taken into account is concerned, the WRF-Chem model has proved to be an excellent candidate for the prediction of the transport and dispersion of natural and anthropogenic aerosols, as well as for volcanic ash and sulfur dioxide emitted during the volcanic eruption. In this study, the output obtained from the numerical model agrees with the satellite data and with the information acquired by the weather stations. It has been possible to reconstruct the dynamics of the Etna eruption  and to confirm the data concerning both the concentration and the direction of propagation of dust and gases emitted into the atmosphere.

More in detail, immediately after the first paroxysm, the ejected material was transported towards the north-east, by winds. Then, the circulation underwent a south-east rotation and, at the end of the eruptive event, the volcanic cloud overlooked the skies of southern Italy, up to the Gargano, the Balkans and the African coasts. Concentrations of SO2 , of the order of 50000 µg m-2 (corresponding to 1.75 Dobson Units), were recorded in the time bands of maximum explosive activity, which progressively decreased to few thousand. The particles expelled during the eruption reached maximum concentrations of 104 µg m-2 and the finest ones (vash_6) went as far as the Balkans and Africa, while those with a larger diameter (vash_4) fell back in the areas surrounding the Etna mountain.

Although the model has exhaustively reproduced the event, it is advisable to modify the WRF-Chem input settings, especially with regard to the particle size distribution to be adopted. The results, obtained by simulating the transport of volcanic ash including particles having dimensions between 15.62 and 31.25 μm (vash_7), suggest the possible spread out of Etna volcanic ash until the east Europe. Upon summarizing the model predictions and the relative comparison with the observed data, it emerges that the WRF-Chem model can be a valid support both to airport management companies affected by volcanic emission phenomena and to other local stakeholders such as public administrations. In addition, the capability of the model to describe the diffusion of Etna volcanic ash and SO2 in the atmosphere is particularly useful in order to better understand the impact of this volcano on the meteorological phenomena involving the whole Mediterranean area. Additional developments of the WRF-Chem model aiming to further increase the accuracy of the predictions are currently underway.

### Author statement

All authors contributed to conceptualization, methodology, investigation, writing, reviewing and editing of the present work.

## Acknowledgments

This work is to be framed within the Progetto di Ricerca e Sviluppo "Impiego di tecnologie, materiali e modelli innovativi in ambito aeronautico (AEROMAT)", Asse II "Sostegno all'innovazione", Area di Specializzazione "Aerospazio" Avviso n. 1735/Ric del 13 luglio 2017 - Codice CUP J66C18000490005, codice identificativo Progetto ARS01_01147 nell'ambito del Programma Operativo Nazionale "Ricerca e Innovazione" 2014-2020 (PON R&I 2014-2020).

Analyses and visualizations used in this paper were produced with the Giovanni online data system, developed and maintained by the NASA GES DISC.

## Data availability

Data can be found at: Physique du Globe de Clermont-Ferrand University: http://wwwobs.univ-bpclermont.fr/SO/televolc/voldorad/; Doppler VOLDORAD 2B: http://voldorad.opgc.fr/home.php; ERA5 – ECMWF: https://cds.climate.copernicus.eu/cdsapp#!/dataset/reanalysis-era5-pressure-levels?tab=overview; NASA EARTH SCIENCE DATA: https://giovanni.gsfc.nasa.gov/giovanni/.

## Declaration of competing interest

The authors declare no competing interest.